# RESULTS OF THE FOUR-WAVE KINETIC INTEGRAL COMPUTATION FOR SPECTRA OF SPECIAL FORMS. THE CASE OF ZAKHAROV SPECTRA.

Vlad Polnikov[1], G.Uma[2]

[1] Obukhov Institute of Atmospheric Physics of RAS, Moscow, Pyzhevskii lane 3,119017 Russia,
Polnikov@mail.ru

[2] Vellore Institute of Technology (VIT University), Chennai, India,
umasathish82@gmail.com

With the aim to show explicitly the non stationarity of the Zakharov spectra, obtained analytically as the stationary solution of the four-wave kinetic equation for stochastic nonlinear surface gravity waves[1,2], we have calculated directly the proper kinetic integral by means of two independent algorithms[3,4] both for an isotropic and anisotropic angular distribution of spectra given on the restricted frequency band.

The four-wave kinetic integral (KI), derived for the first time by K. Hasselmann[5], has the form

$$\frac{\partial N(\mathbf{k})}{\partial t} = \iiint T(\mathbf{k},\mathbf{k}_1,\mathbf{k}_2,\mathbf{k}_3)\delta(\mathbf{k}+\mathbf{k}_1-\mathbf{k}_2-\mathbf{k}_3)\delta(\omega+\omega_1-\omega_2-\omega_3)\times \\ \times\{N_2 N_3(N+N_1)-N_1 N(N_2+N_3)\}d\mathbf{k}_1 d\mathbf{k}_2 d\mathbf{k}_3 \quad (1)$$

where $N_i = N(\mathbf{k}_i)$ is the wave action spectrum, related to the wave energy spectrum $S(\mathbf{k}_i) \propto \omega N(\mathbf{k}_i)/g$, $T(\mathbf{k},\mathbf{k}_1,\mathbf{k}_2,\mathbf{k}_3)$ is the matrix element for four-wave interactions, $\delta(\mathbf{k})$ и $\delta(\omega)$ are the Dirac δ-functions describing resonance feature of the interactions, and frequency ω is related to wave vector **k** by the dispersion relation $\omega^2 = gk$.

Zakharov and coauthors [1,2] have analytically found on the infinite frequency band and for the case of isotropic angular distribution of the frequency-angular spectrum $S(\omega,\theta)$ the stationary solution of (1). These solutions have the form

$$S_E(\omega) = c_1 P_E^{1/3} g^{4/3} \omega^{-4} \propto \omega^{-4} \quad (2a)$$

$$S_N(\omega) = c_2 P_N^{1/3} g \omega^{-11/3} \propto \omega^{-11/3} \quad (2b)$$

which are treated as the Kolmogorov type spectra with the constant wave energy flux to high frequencies (flux up), $P_E$, and constant wave action flux down, $P_N$ ($g$ is the gravity acceleration).

Here we show explicitly that such a kind spectra are not making kinetic integral (1) equal to zero on the finite frequency band both in the isotropic and anisotropic angular distribution for $S(\omega,\theta)$.

To this aim, we use two independent algorithms of calculating KI [3,4]. As the integrand we use the frequency-angular spectrum $S(\omega,\theta)$ of the form

$$S(\omega, \theta) = \left[\frac{\exp(-(X/4)\cdot(\omega/\omega_p)^{-4})}{\omega^X \exp(-X/4)\omega_p^{-X}}\right] \cos^n(\theta/2) \sim \omega^{-X}. \tag{3}$$

Here $X$ is the power of the failing law of the spectrum, the multiplier in square brackets is the cutting factor $M(\omega,\omega_p)$ providing the peak of spectrum at the peak frequency $\omega_p$, and $n$ is the power of the simplest angular function of the spectrum, the shape of which is not principal here. The spectrum is normalized so that $S(\omega_p,0) \equiv S_p = 1$ at $\omega_p$. For $n = 0$ we have isotropic spectrum in angle, and varying $n$ allows to expand the task for anisotropic case.

The choice of calculating grid $\{\omega, \theta\}$ is very important for accuracy of the calculation result. After some attempts, we have chosen the grid

$$\omega(i) = \omega_0 q^{i-1} \quad (1 \le i \le N); \quad \theta(j) = -\pi + (j-1)\cdot\Delta\theta \quad (1 \le j \le M); \tag{4}$$

with parameters: $\omega_0 = 0.5568374$, $q = 1.05$, $N = 61$, $\Delta\theta = 2\pi/M$, and $M = 18$. In such a case $\omega_p = 1$, and the highest frequency is $\omega_m = \omega(61) = 10.4$. The latter provides a relatively great frequency band, where the calculating results for KI do not depend on the choice for $\omega_0$ and $\omega_m$. Here we should mention that the converging of KI for the spectra of the form (2) is relatively bad, as far as matrix function $T(\mathbf{k},\mathbf{k}_1,\mathbf{k}_2,\mathbf{k}_3)$ is growing as $\omega^{12}$. This circumstance make calculating KI for the Zakharov spectra a very delicate scientific problem.

According to [3], the nonlinear transfer function, $Nl(\omega,\theta) = \partial S(\omega,\theta)/\partial t$, following from KI calculation, can be expressed in the form

$$Nl(\omega,\theta) = g^{-4} S_p^3 \omega_p^{11} \Phi(\omega,\theta)_{nl} \tag{5}$$

where in dimensionless function $\Phi(\omega,\theta)_{nl}$ is to be analyzed here. After integration $\Phi(\omega,\theta)_{nl}$ over angles, one-dimensional function $Nl(f)$ will be shown on the dimensionless frequency axis, $f = \omega/\omega_p$, with no regard to values of spectral peak and exact dimensional values of $Nl(f)$.

Results are shown below for the restricted frequency band ($\omega_0 < \omega/\omega_p < 6$) which is most important from the practical point of view.

**Result found by Algorithm** [3]

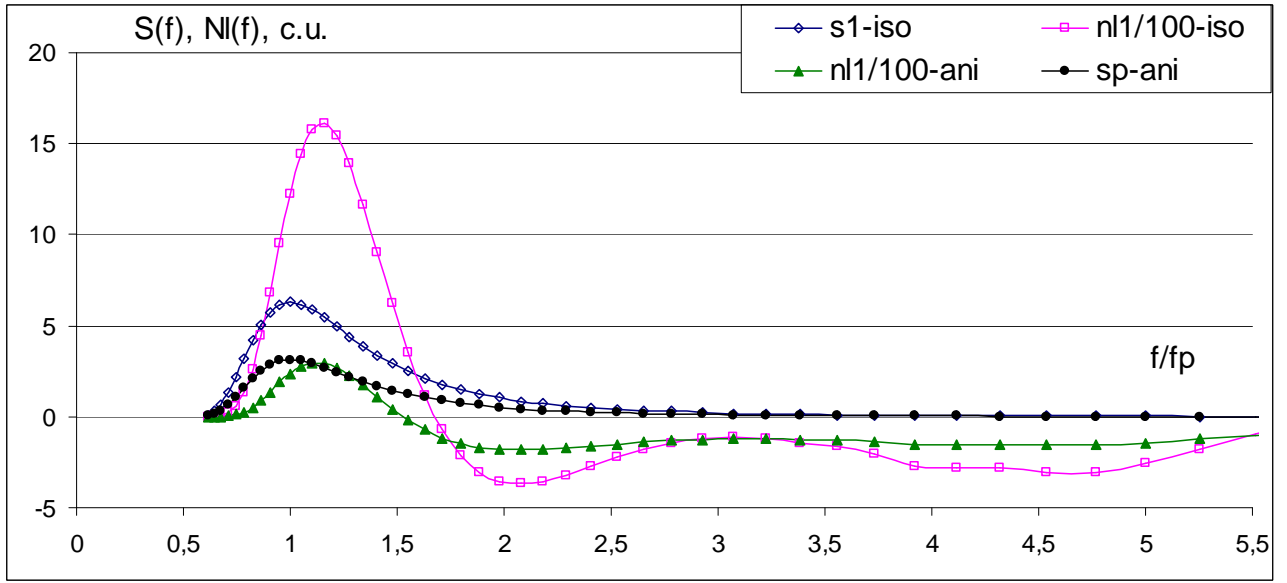

Fig1. 1D spectrum *S1(f)* and nonlinear transfer of energy *Nl1(f)*
for 2 types of Zakharov spectrum with X = 4.
"Iso" means n = 0. "Ani" means n = 2.

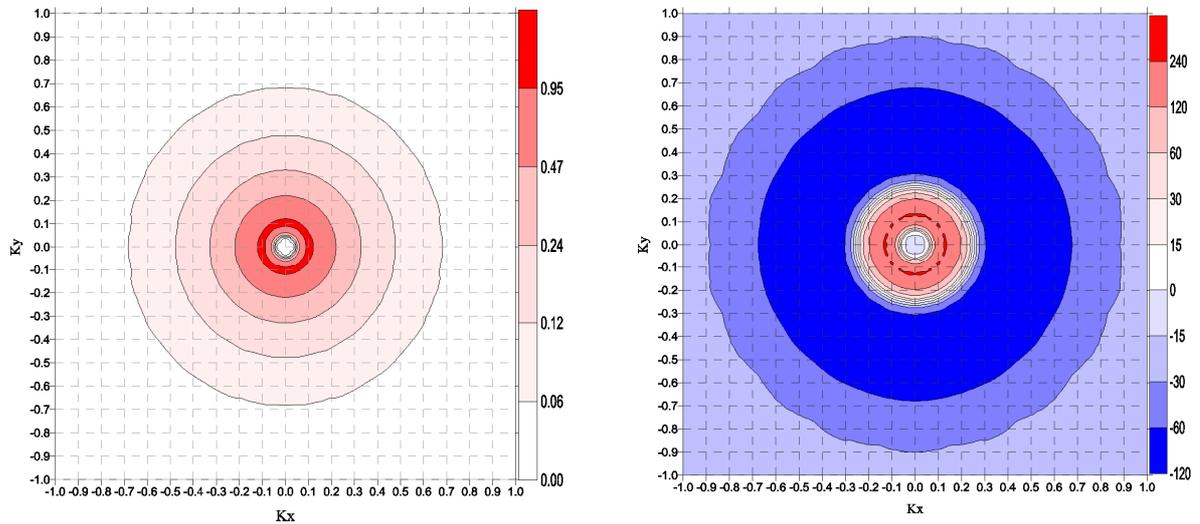

Fig. 2. 2D spectrum $S(\omega,\theta)$ and nonlinear transfer of energy $\Phi(\omega,\theta)_{nl}$ in the **k**-space
for isotropic Zakharov spectrum (X=4, *n* = 0)

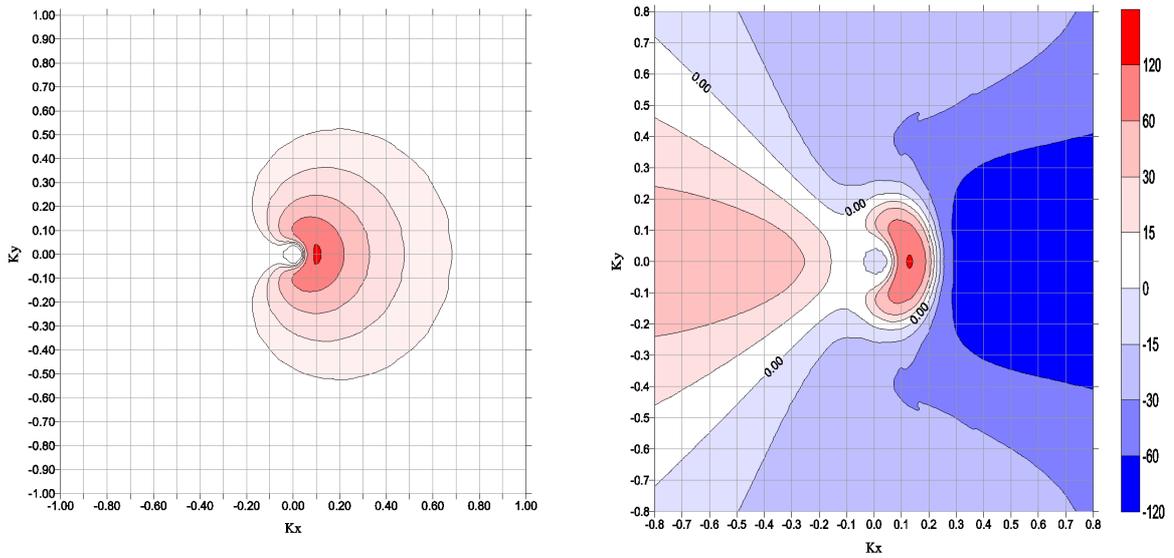

Fig. 3. 2D spectrum $S(\omega,\theta)$ and nonlinear transfer of energy $\Phi(\omega,\theta)_{nl}$ in the **k**-space for anisotropic Zakharov spectrum (X=4, $n=2$)

**For the aim of comparison between specific and standard spectra**
(evidence of algorithm quality)

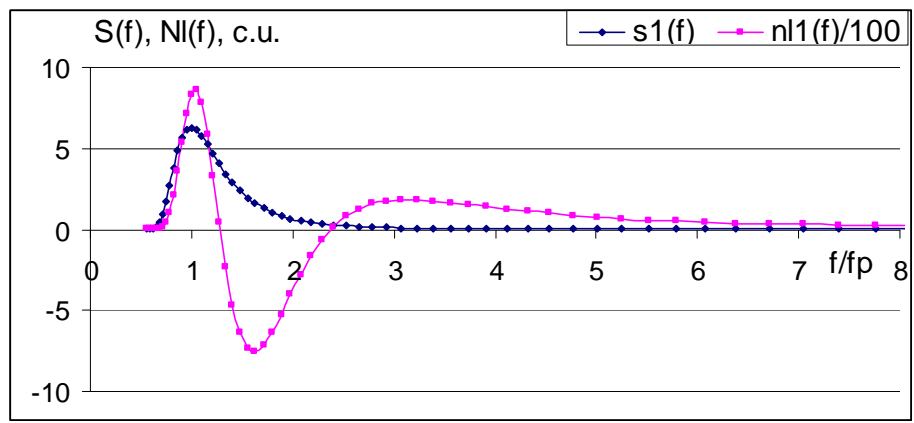

Fig. 4. 1D-spectrum and nl-transfer of energy for isotropic Pierson-Moskowitz spectrum (X = 5)

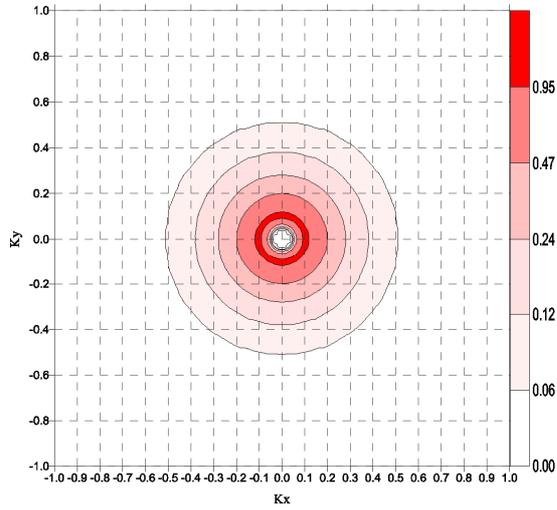 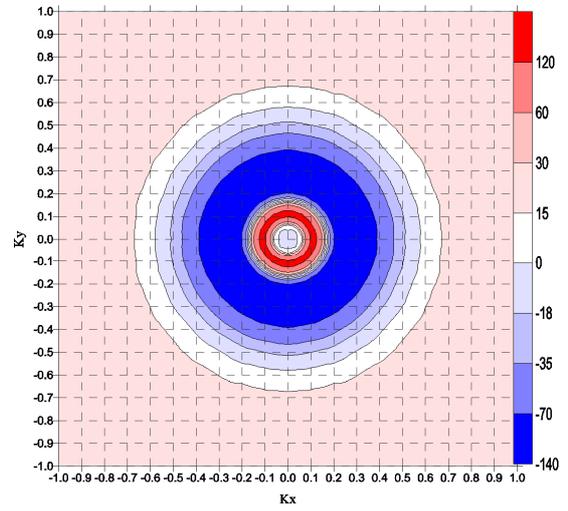

a) isotropic case

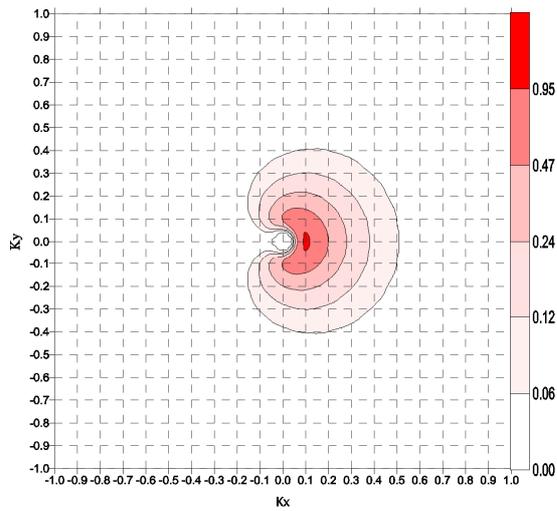 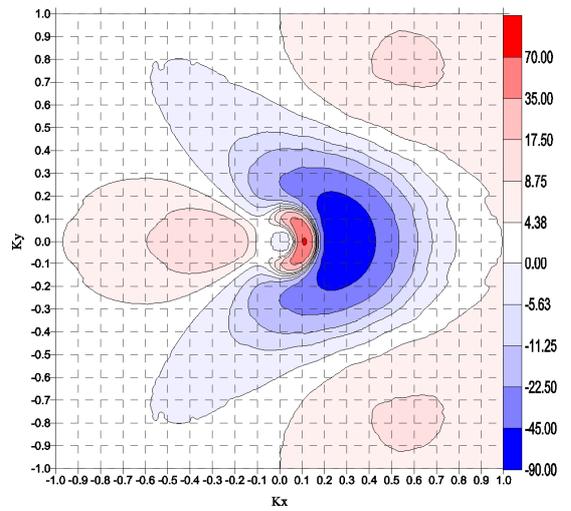

b) anisotropic case

Fig. 5. The same as in Figs 2 and 3 but for Pierson-Moskowitz spectrum

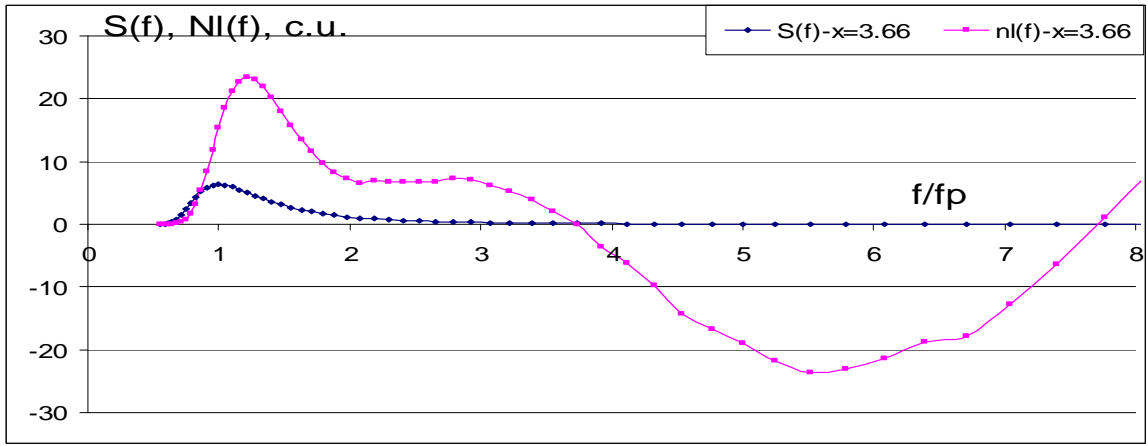

Fig. 6. 1D spectrum S(f) and nl-transfer of energy *N1*(*f*)
for isotropic Zakharov-Zaslavskii spectrum (X = 11/3, n = 0).

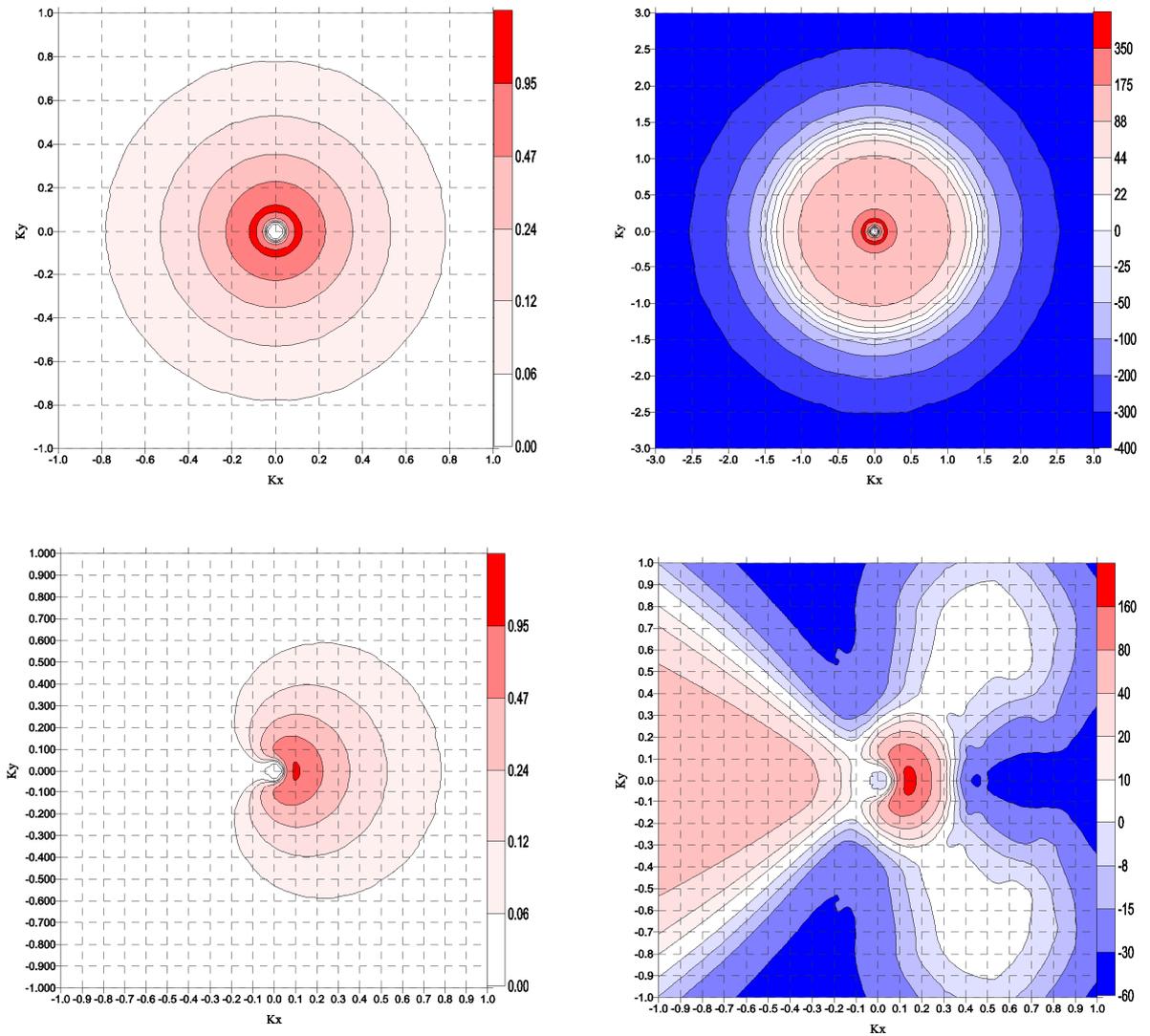

Fig. 7. The same as in Figs 2 and 3 but for Zakharov-Zaslavskii spectrum (n = 11/3, n = 2)

**Results found by Algorithm [4]**

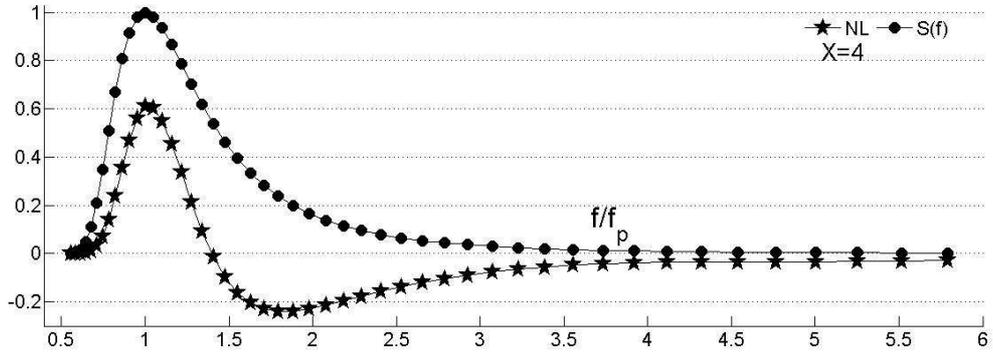

a)

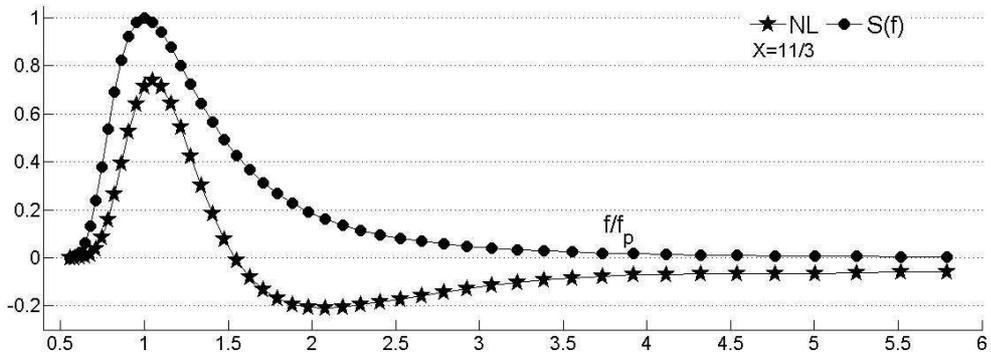

b)

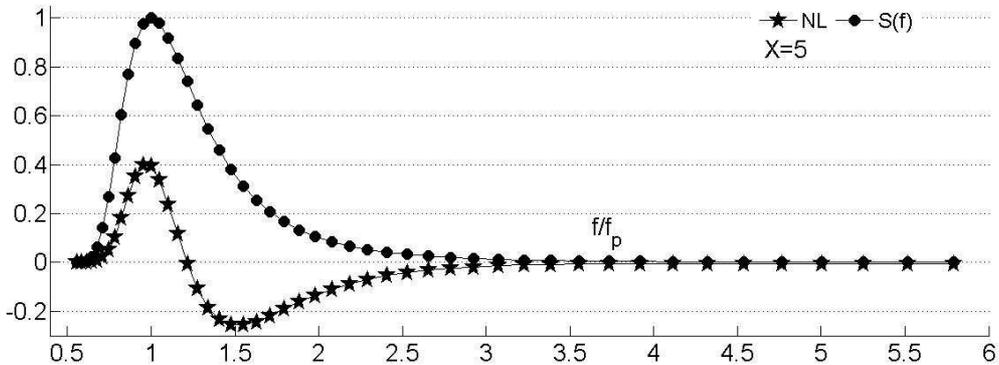

c)

Fig. 8. 1D spectra S(f) and nl-transfer of energy *N1(f)*
for 2 types of Zakharov spectrum ( X = 4, X= 11/3)
and Pierson-Moskowitz spectrum.